\documentstyle[12pt]{article}
\input epsf.tex
\newlength{\halfpage}
\begin{document}

\raggedright

\begin{center}
{\large \bf  EJECTION OF MATTER AND ENERGY FROM NGC 4258\\}
{\ \\}
{\bf E.M. Burbidge and G. Burbidge\\}
{\bf Department of Physics and Center for Astrophysics \& Space 
Sciences\\}
{\bf University of California, San Diego\\}
{\bf La Jolla, California 92093-0111, USA}
\end{center}

\setcounter{page}{1}
\begin{center}
{\bf ABSTRACT}
\end{center}

It has been claimed that the megamaser observations of the nucleus of 
NGC 4258 show that a massive black hole is present in its center 
(Miyoshi et al. 1995, Greenhill et al. 1995).  We show that the 
evidence of ejection of gas, radio plasma, and X-ray emitting QSOs from 
this nucleus all show that the ejection is coming from the center 
in a curving flow within 
a cone with angle  $\sim 40^{\circ}$, centered at P.A. $100^{\circ}$.  
This is close to the direction 
in which the velocities from the megamaser have been measured, so that 
the evidence {\em taken as a whole}  suggests that the masering gas 
also is being {\em ejected} in the same direction at velocities $ \pm 
900$
km sec$^{-1}$ and not rotating about a massive black hole.  Thus it does 
not provide evidence for a black hole in the center.

\noindent {\em Subject headings:} galaxies: nuclei: individual (NGC 4258) -- black 
holes -- masers


\begin{center}
{\bf 1. INTRODUCTION}
\end{center}

There are no new observations in this paper.  Its purpose is to 
summarize evidence collected since 1962 from 
the measurements of ionized gas, continuum radio observations, and 
X-rays, all of which shows that violent activity taking place close to the 
center of NGC 4258 is giving rise to ejection of gas, relativistic 
particles and coherent objects and to show that these observations 
together with 
the megamaser observations all suggest that matter is being ejected from 
the center.  However, contrary to what has been frequently claimed from 
the megamaser observations alone (Miyoshi et al. 1995), they do not 
provide evidence for the existence of a massive black hole.

\begin{center}
{\bf II.  OPTICAL, RADIO AND X-RAY OBSERVATIONS OF NGC 4258}
\end{center}

NGC 4258 is one of the nearest galaxies with an active nucleus.  Its 
distance $D \approx 7 Mpc$, although some investigators have used a 
smaller value.  Its systemic velocity $V_{\circ} = 465 $ km  s$^{-1}$. 
The low-ionization strong emission lines in the nucleus lead to a 
classification as a 
LINER galaxy rather than as a Seyfert galaxy, although it was in Seyfert's 
original list (cf Burbidge \& Burbidge 1962).

A study of the velocity field in the ionized gas in the bright knotty 
spiral arms was made as part of our original program of measurement of rotation 
curves of spiral galaxies (Burbidge, Burbidge \& Prendergast 1963).
This showed 
the galaxy is quite massive ($\sim 10^{11} M_{\odot}$ interior to our 
last measured point), 
but it also showed that substantial non-circular motions were present 
within 40 arcsec of the apparent nucleus.  In that study, a position 
angle of $157^{\circ}$ was taken to be the direction of the major axis.  
The ``non-circular'' motion occurred in P.A. $157^{\circ}$, 
$167-169^{\circ}$, and $126^{\circ}$ (see Fig 1 for these directions).  
A formal analysis of these 
velocities in terms of the mass model of similar spheroidal 
surfaces, balancing rotation against the gravitational pull of the mass 
interior to the measured points, gave, for any choice of axial ratio, 
the unphysical result of {\em negative} densities at $\sim$ 1 arc min 
from the center.  Burbidge et al. hence concluded that radial 
motions of $\sim 300$ km  s$^{-1}$ were present, and suggested that 
the mass distribution of NGC 4258 might resemble a barred spiral.

\begin{figure}[h]
\epsfbox{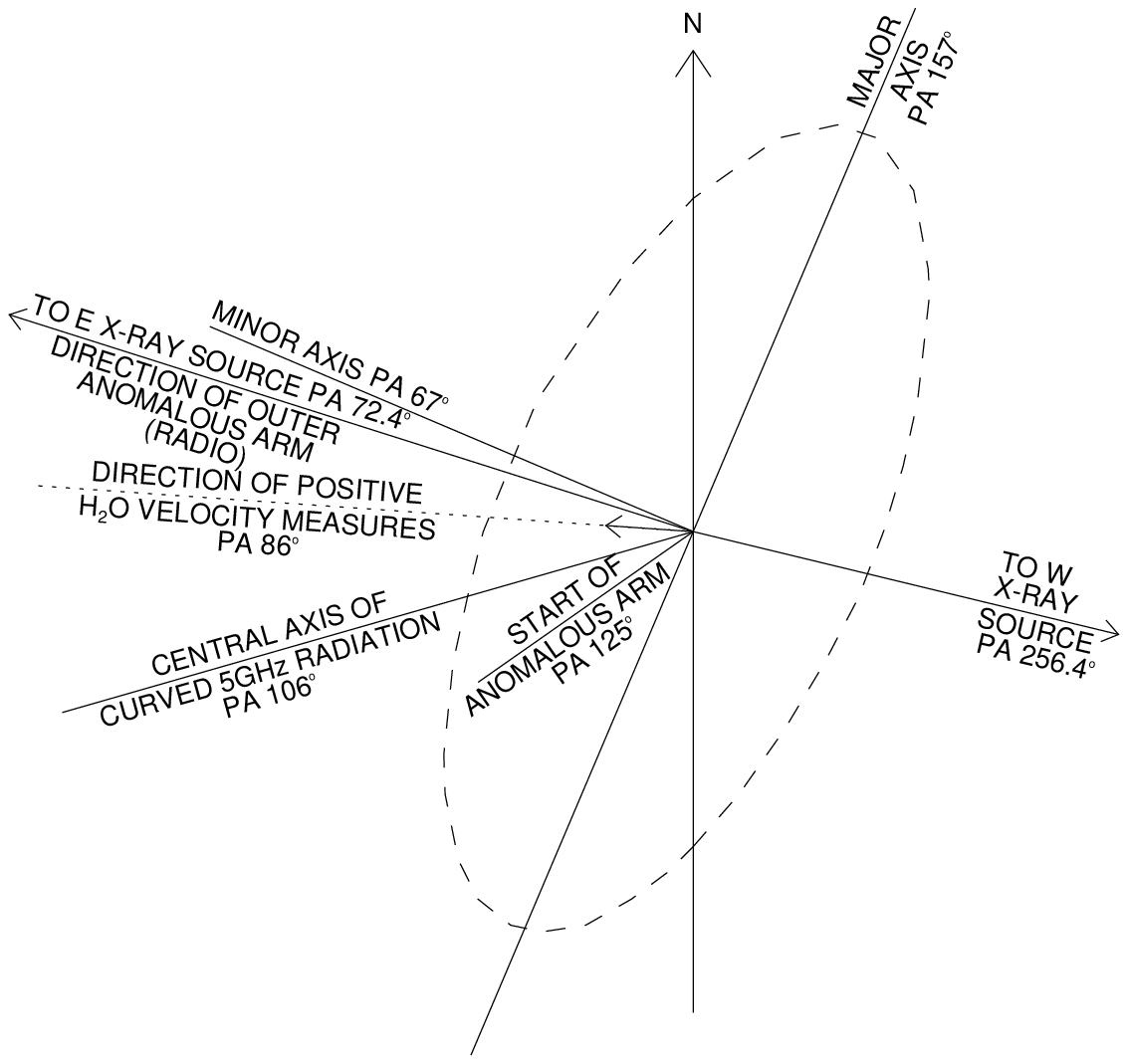}
\caption{Schematic diagram of NGC 4258 showing position angles of major 
and minor axes, and various directions in the plane of the sky of 
apparent ejection from the center (ionized gas, relativistic particles 
for radio emission, X-ray sources).}
\end{figure}

The existence of non-circular motions in NGC 4258 was confirmed by 
several later studies (cf. van der Kruit 1974), and they were 
linked to the discovery by Court\`{e}s and Cruvellier (1961) of a pair of 
``anomalous arms'' in P.A. $125^{\circ}$. Optically, these are  visible in 
H$\alpha$ emission 
only.  Since these arms have no stellar continuum from stars expected to be 
present  
to ionize the hydrogen (Court\`{e}s et al. 1993), it was concluded that 
these ``arms'' are not normal quasi-stationary 
density concentrations, but hot gas which is being ejected.

Van der Kruit, Oort \& Mathewson (1972) measured the radio continuum at 
1415 MHz, and found smooth curved ridges of continuum which differ in 
position, shape, and continuity from  the normal optical arms, but  
close to the center they coincide with the filamentary H$\alpha$ 
arms of Court\`{e}s \& Cruvellier.  This study of the optical and radio 
data led van der Kruit et al.
 to the conclusion that there had been ejection from the 
nucleus of clouds in two opposite directions about $18 \times 10^{6}$ 
years ago, at velocities ranging from 800 to 1600 km s$^{-1}$.

High resolution radio continuum observations made at 1480 MHz by van 
Albada \& van der Hulst (1982) showed that the radio emission that lies 
along the SE anomalous arm curves around sharply eastward and ends in a 
tail directed toward the NE.  Even more striking detail is shown in Fig 
2c of the paper by Cecil et al. (1995), 
which is based on measurements at 5 GHz.  The position angle of the 
central axis of this emission near the center, 106$^{\circ}$, is shown 
in Fig 1.

Cecil et al. (1995) also used data from Ford et al. (1986), who 
discovered that the Court\`{e}s \& Cruvellier anomalous arms or jets
were resolved into distinct helical strands braided around each other, 
suggesting the influence of a magnetic field in the center reacting with 
the outward flow of the plasma.  A combination of X-ray data with the optical 
and radio data led Cecil et al. to a model in which hot shocked gas at 
500 km s$^{-1}$ might have been entrained as the outward-moving 
nuclear jets reacted with the molecular clouds of CO that had been 
mapped as lying closely along the anomalous arms (Krause et al. 
1990).  The strength of a magnetic field along the anomalous arms is 
indicated by the high degree of polarization in the radio emission at 
1.49 GHz measured by Hummel et al. (1989).  Their Fig 3 shows this 
very clearly, and they point out that the magnetic field configuration 
is bisymmetric, with the field lines closely following the ridges of the 
radio emission.

A detailed study of the emission-line velocity field in the central few 
arc min, by Rubin \& Graham (1990) using the KPNO echelle spectrograph
on the 4-m Mayall telescope, shows the complexity of the ionized gas flow 
close to the nucleus.  Their plots of velocity in PA $150^{\circ}$ and 
$125^{\circ}$ reveal multiple-valued velocities, and they described the 
velocity details as ``more spectacular than those we have seen in any 
other galaxy''.
Rubin \& Graham suggested that this indicated that these are due to 
jets bursting out of the galactic plane, with 
gas drizzling back into the plane at positions indicated on their 
velocity plots.

All of these studies are best explained by high velocity jets 
emanating from the nucleus, including (or being initiated by) 
relativistic electron 
(or electron-positron) jets emitting synchrotron radiation, as indicated 
by the radio data.  

ROSAT PSPC X-ray observations of NGC 4258 and its vicinity (Pietsch et 
al. 1994) showed the anomalous arms or jets in X-rays for the first 
time, most clearly in the hard (H2) band where absorption due to the 
disk of the galaxy is minimum.  These data also showed two bright 
compact X-ray sources, each $\sim$9 arc min from the center of the 
galaxy, the E source in P.A. $72.^{\circ}4$ with respect to the center 
and the W source in P.A. 
$256 \! \stackrel{\circ}{.} \! 4$; these directions are shown in Fig 
1.  The 
line joining the two sources passes close to the 
center and lies within some $5^{\circ}$ of the minor axis.  An overlay 
of the directions to compact external sources on the 1480 MHz radio map 
of van Albada \& van der Hulst (Fig 8 of Pietsch et al.) shows this 
quite clearly, and Pietsch et al. point out that the line connecting the 
compact sources touches the ends of the anomalous arms.  Following this 
suggestion by Pietsch et al. 
that the X-ray compact sources might have been ejected from the 
nucleus of NGC 4258, Burbidge (1995) obtained spectra of them and showed 
that they are indeed a pair of QSOs with redshifts $z = 0.398$
 (W source) and 0.653 
(E source), and discussed the implications if these have been ejected 
from the nucleus of NGC 4258.

\begin{center}
{\bf 3.  MEGAMASER OBSERVATIONS AND THEIR INTERPRETATIONS}
\end{center}

We turn now to the detection of an H$_{2}$O megamaser in the nucleus
of NGC 4258.  This was reported by Claussen et al. (1984), and Miyoshi et al. 
(1995) detected maser emission at velocities offset by 
$\pm 750$ to 
1000 km s$^{-1}$, bracketing the emission at the systemic velocity
of $\pm 465$ km s$^{-1}$.  They have concluded that relative to the 
center the velocities of the excited gas cover the range 
$\pm 900$ km 
s$^{-1}$.  The figure of their model shows blobs spaced almost 
equidistantly on the high-velocity side.
According to their model, the velocities and spacing relative 
to the nucleus of NGC 4258 are perpendicular to the line of sight, and 
 arise in a disk oriented in P.A. 
86$^{\circ}$, at an angle $\sim 11^{\circ}$ to the minor axis of NGC 4258 as 
estimated from its image (Cecil et al. 1995) (see Fig 1).  
They quoted  Hubble (1943) 
who had determined from the dark lanes that the spiral arms are trailing.  
Miyoshi et al. give 
their determination of the angle between the spin axis of this H$_{2}$O 
megamaser disk and the spin axis of the galaxy as $119^{\circ}$.  
Miyoshi et al. chose to interpret their velocity observations as being 
due to circular motion involving a disk surrounding a massive black 
hole.  Thus they have claimed that their observations provide direct 
evidence for the existence of a massive black hole in the center of NGC 
4258.  

Now it is well known that velocity shifts alone 
detected in the central region of a galaxy may be due to rotation, or to 
radial motion inward or outward, or to a combination of rotation and radial 
motion.  Since one is trying to construct a 3-dimensional model from 
2-dimensional 
data, without an {\em independent} measurement of 
the position angle of the line of nodes, or velocity measurements in many 
different position angles, it is impossible to determine whether the 
motions are purely circular or radial or both.
However, what has been done by Miyoshi et al. is to {\em assume
without proof} that they are looking at circular motion about a black 
hole -- i.e. they 
assume that the black hole-accretion disk 
paradigm is correct, and in so doing determine 
the position angle of the rotating disk as $86^{\circ}$.
In this way they are 
able to calculate a central mass, and this is being widely claimed to be 
observational evidence for the presence of a massive black hole.  For other 
galaxies with megamasers in which no motions have been measured, 
reference is made to NGC 4258 as {\em the} evidence for a black hole.  Thus 
the idea is growing up that the existence of megamasers is a further 
indication of the existence of massive black holes in active galaxies.  
This is a good example of the disservice that {\em belief} in the paradigm is 
doing to a true understanding of the activity in galaxies.

However, the fact is that the direction of the velocity vectors directly 
observed from the 
maser line components is close to that for optical, radio and X-ray 
ejections (Arp 1996). This is shown in Fig 1.  
Thus the {\em simplest explanation} devoid of any 
further assumption, is that the gas excited in the masering process is being 
ejected by the same unknown process and in the same direction as all of 
the matter and energy further out.  This may be connected with the 
linear drift of 9.5 km s$^{-1}$yr$^{-1}$ in those features with 
velocities $\pm$150 km s$^{-1}$ of the systemic velocity, while the 
overall velocity range remains stationary over time (Greenhill et al. 
1995a).

\begin{center}
{\bf 4. CONCLUSION}
\end{center}

As we have shown all of the evidence from 
optical, radio continuum and X-ray studies suggests that explosive events 
in the nucleus have given rise to this activity.  Whatever the process 
is, it has restricted the ejection to a cone of angle $\sim 40^{\circ}$ 
containing the minor axis of NGC 4258, and approximately centered on the 
minor axis.
The megamaser observations fit very well into this picture, since the 
P.A. of the velocities,  $86^{\circ}$, is only $\sim 11^{\circ}$ from the 
direction of the minor axis.

Thus the observations taken as a whole 
point towards ejection in  the same cone,
not rotation about a black hole.  By arguing that the megamaser 
observations are to be interpreted as evidence of rotation, Myoshi et al. 
have to arrange that the rotation axis of the purported disk is oriented 
practically at right angles to the direction of ejection of the other 
material.

No one knows what the primary source of energy is.  It may be 
gravitational energy released in the collective evolution of a dense 
star cluster, or a massive black hole, or the creation of matter at the 
nucleus.  While the most popular model is based on the release of 
gravitational energy in the black hole-accretion disk model (cf Rees 
1984), and this has turned into the black hole-accretion disk paradigm, 
the observations do not necessarily indicate that this is correct.
Unfortunately the general tendency of observers 
has been to {\em interpret} the 
observations in terms of this model rather than testing it (cf 
Holt et al. 1992).
The simplest model required to explain all of the observations of all of 
the activity in the nucleus of NGC 4258 must be able 
\begin{enumerate}
\item  To provide 
energy to excite and maintain what appear to be a number of regularly 
spaced gas clouds as is indicated by the maser observations.
\item To provide energy to eject gas at high speed ($\sim 1000$ km 
s$^{-1}$) over time scales of $\sim 10^{6} - 10^{7}$ years or longer.
\item To provide energy that will generate radio synchrotron emission 
giving rise to the complex radio arms.
\item To eject compact X-ray sources each centered on QSOs with 
anomalous redshifts.
\end{enumerate}

There is certainly a large mass concentration in the central region of 
NGC 4258, but evidence for a black hole is only 
obtained by interpreting the megamaser observations in a special way, and 
ignoring the other evidence.  Optical spectroscopy with high spatial 
resolution in the central 10-20 arc sec may help to define the properties 
of the nuclear region.

This research was supported in part by NASA grant NAG5-1630. 

\pagebreak

\begin{center}
{\bf REFERENCES}
\end{center}

\begin{description}
\item[] Arp, H.C. 1996, private communication
\item[] Burbidge, E.M. 1995, A\&A, 298, L1
\item[] Burbidge, E.M. \& Burbidge, G.R. 1962, ApJ, 135, 694
\item[] Burbidge, E.M., Burbidge, G.R. \& Prendergast, K.H. 1963, ApJ, 
138, 375
\item[] Cecil, G., Wilson, A.S. \& dePree, C. 1995, ApJ, 440, 181
\item[] Claussen, M., Heiligman, G., Lo, K.Y. 1984, Nature, 310, 298.
\item[] Court\`{e}s, G. \& Cruvellier, P. 1961, C. R. Acad. Sci. Paris, 253, 
218
\item[] Court\`{e}s, G., Petit, H., Hua, C.T., Martin, P., Blecha, A., 
Huguenin, D. \& Golay, M. 1993, A\&A, 268, 419
\item[] Ford, H., Dahari, O., Jacoby, G.H., Crane, P.C. \& Ciardullo, R. 
1986, ApJ, 311, L7
\item[] Greenhill, L.J., Jiang, D.R., Moran, J., Reid, M., Lo, K.Y. \& 
Claussen, M.J. 1995, ApJ, 440, L19
\item[] Greenhill, L.J., Henkel, C., Becker, R., Wilson, T.L. \& 
Wouterloot, J.G.A. 1995a, A\&A, 304, 21
\item[] Holt, S.S., Neff, S.G. \& Urry, C.M. 1992 ``Testing the AGM 
Paradigm'', AIP Conf. Proc. 254 
\item[] Hubble, E. 1943, ApJ, 97, 112
\item[] Hummel, E., Krause, M. \& Lesch, H. 1989, A\&A, 211, 266
\item[] Krause, M., Cox, P., Garcia-Barreto, J.A., Downes, D. 1990, 
A\&A, 233, L1
\item[] Miyoshi, M., Moran, J., Herrstein, J., Greenhill, L., Nakal, N., 
Diamond, P., Inone, M. 1995, Nature, 373, 127
\item[] Nakai, N., Inone, M. \& Miyoshi, M. 1993, Nature, 361, 45
\item[] Pietsch, W., Vogler, A., Kahabka, P., Jain, A., \& Klein, V. 
1994, A\&A, 284, 386
\item[] Rees, M.J. 1984, ARA\&A, 32, 471
\item[] Rubin, V. \& Graham, J. 1990, ApJ, 362, L5
\item[] van Albada, G.D. \& van der Hulst, J.M., 1982, A\&A, 115, 263
\item[] van der Kruit, P.C. 1974, ApJ, 192, 1
\item[] van der Kruit, P.C., Oort, J.H. \& Mathewson, D.S. 1972, A\&A, 
21, 169
\end{description}

\end{document}